
\input phyzzx
\hoffset=0.2truein
\voffset=0.1truein
\hsize=6truein
\def\TITLEPAGE{\frontpagetrue}
\def\CALT#1{\hbox to\hsize{\tenpoint \baselineskip=12pt
        \hfil\vtop{
        \hbox{\strut CALT-68-#1}}}}

\def\CALTECH{
        \address{California Institute of Technology,
Pasadena, CA 91125}}

\def\AUTHOR#1{\vskip .2in \centerline{#1}}
\def\ANDAUTHOR#1{\smallskip \centerline{\it and} \smallskip
\centerline{#1}}
\def\ABSTRACT#1{\vskip .2in \vfil \centerline{\twelvepoint
\bf Abstract}
        #1 \vfil}
\def\ENDTITLEPAGE{\vfil\eject\pageno=1}

\tolerance=10000
\hfuzz=5pt
\tolerance=10000
\hfuzz=5pt

\def\>{{\buildrel >\over {{}_{\sim}}}}
\def\<{{\buildrel <\over {{}_{\sim}}}}
\def\-{{\buildrel <\over {{}_{-}}}}

\TITLEPAGE\CALT{1911}         
\bigskip
\titlestyle {Long Distance Contribution to $K^+ \rightarrow \pi^+ \nu
\bar\nu$\foot{Work supported in part by the U.S. Dept. of Energy
under Grant no. DE-FG03-92-ER 40701.}}
\AUTHOR{Ming Lu}
\ANDAUTHOR{Mark B. Wise}
\CALTECH

\ABSTRACT{We estimate the long distance contribution to the $K^+ \rightarrow
\pi^+ \nu \bar\nu$ decay amplitude using chiral perturbation theory.  We find
that it is likely to be less than a few percent of the part of the short
distance contribution proportional to the square of the charm quark mass.  In
the large $N_c$ limit the leading (in chiral perturbation theory) long distance
contribution to $K^+ \rightarrow \pi^+ \nu \bar\nu$ from Feynman diagrams with
a $W$ and $Z^0$ vanishes.}
\ENDTITLEPAGE
\eject

The decay $K^+ \rightarrow \pi^+ \nu \bar\nu$ is likely to be observed in the
 near future.  An accurate measurement of  its
 branching ratio can provide a precise determination of the weak mixing angle
 $V_{td}$ (once the $t$-quark mass is known).  The general form for the $K^+
\rightarrow \pi^+ \nu \bar\nu$
 invariant matrix
 element is
$$	{\cal M} (K^+ \rightarrow \pi^+ \nu \bar\nu) = {G_F\over \sqrt{2}} {\alpha
 f_+\over 2\pi \sin^2 \theta_W} \Bigg[V_{ts}^* V_{td} \xi_t (m_t^2/M_W^2)$$
$$	V_{cs}^* V_{cd} \xi_c (m_c^2/M_W^2) + V_{us}^* V_{ud} \xi_{LD}\Bigg]
 (p_K + p_\pi)^\mu \bar u (p_{\nu}) \gamma_\mu (1 - \gamma_5)v
(p_{\bar\nu})\,\,
. \eqno (1)$$
In eq. (1) $G_F$ is the Fermi constant, $\theta_W$ is the weak mixing angle,
 $\alpha$ is the fine structure constant, $f_+$ is the form factor in $\bar K^0
\rightarrow \pi^+ e \bar\nu_e$ decay and $V_{ab}$ denotes the $a \rightarrow b$
 element of the Cabibbo-Kobayashi-Maskawa matrix.  The factors $\xi_t$ and
$\xi_c$ arise from the short distance contribution to $K^+ \rightarrow \pi^+
\nu \bar\nu$ and $\xi_{LD}$ from the long distance contribution.  Neglecting
perturbative strong interaction corrections [1]
$$	\xi_t (x) = {x\over 8} \left[ - {2  + x\over 1 - x} + {3x-6\over (1 -
x)^2}  \ln x\right] \,\, , \eqno (2a)$$
$$	\xi_c (x) \simeq {x\over 8} \left[-6 \ln x - 2\right]\,\, . \eqno
(2b)$$
Recently QCD corrections to $\xi_t (x)$ of order $\alpha_s(m_t)$ have been
 calculated [2] and the QCD corrections to $\xi_c (x)$ have been summed in the
next to leading logarithmic approximation [2,3].  The value of $\xi_c$ is about
 $10^{-3}$ . (This is for $\nu_e$ and $\nu_{\mu}$ neutrinos.
  A somewhat
 smaller value is obtained for $\nu_{\tau}$ because the $\tau$ lepton mass
cannot
 be neglected in the $W$-box diagram.)
 With the next to leading logarithms included the largest uncertainty in
 $\xi_c$ comes
from our imprecise knowledge of the charm quark mass and $\Lambda_{QCD}$  .

This paper contains an estimate of $\xi_{LD}$ using chiral perturbation theory.
  Previous estimates of this type were made by Rein and Sehgal [4] and Hagelin
and Littenberg [5].  Our work is similar to theirs in approach and conclusions,
 however, some of the details are different.

One of the most prominent features of the pattern of kaon (and hyperon)
decays is the $\Delta I=1/2$ rule. Nonleptonic kaon decay amplitudes
that arise from the $I=1/2$ part of the $\Delta S=1$ effective weak
Hamiltonian are
enhanced by a factor of twenty over those that arise from the $I=3/2$
part. In this letter we focus primarily on the part of $\xi_{LD}$ that
arises from the time ordered product of the weak $\Delta S=1$ effective
Hamiltonian
with the $Z^0$ neutral current, since it receives a $\Delta I=1/2$
enhancement. The $Z^0$ coupling to light $u$, $d$, and $s$ quarks is given
by
     $${\cal L}_{int} = {\sqrt{g_1^2+g_2^2}\over
2}Z^{0\mu}~[J_{\mu}^{(L)}-2~{\rm
 sin^2\theta_W}J_{\mu}^{(e.m.)}]\,\, , \eqno (3)$$
where $J_{\mu}^{(L)}$ is the left handed current
$$ J_{\mu}^{(L)}=\bar u_L\gamma_{\mu}u_L-\bar d_L\gamma_{\mu}d_L-
\bar s_L\gamma_{\mu}s_L \,\, , \eqno (4)$$
(the part of $J_\mu^{(L)}$ involving the strange quark was neglected in Ref.
[4]) and $J_{\mu}^{(e.m.)}$ is the electromagnetic current
$$J_{\mu}^{(e.m.)}={2\over 3}\bar u\gamma_{\mu}u-{1\over 3}\bar d\gamma_{\mu}
d-{1\over 3}\bar s\gamma_{\mu}s \,\, . \eqno(5)$$
The left-handed current $J_{\mu}^{(L)}$ can be written as the sum of a
piece that transforms as $(8_L,1_R)$ and a piece that transforms as $(1_L,1_R)$
with respect to the chiral symmetry group $SU(3)_L\times SU(3)_R$.
$$ J_{\mu}^{(L)}=J_{8\mu}^{(L)}+J_{1\mu}^{(L)}\,\, , \eqno (6)$$
where
$$J_{8\mu}^{(L)}={4\over 3}\bar u_L\gamma_{\mu} u_L-{2\over 3}\bar
  d_L\gamma_{\mu} d_L-{2\over 3}\bar s_L\gamma_{\mu} s_L\,\, , \eqno (7)$$
$$J_{1\mu}^{(L)}=-{1\over 3}\bar u_L\gamma_{\mu} u_L-{1\over 3}\bar
  d_L\gamma_{\mu} d_L-{1\over 3}\bar s_L\gamma_{\mu} s_L\,\, . \eqno (8)$$
The electromangnetic current transforms as $(8_L,1_R)+(1_L,8_R)$ under
$SU(3)_L\times SU(3)_R$.

The interactions of kaons, pions, and the eta are constrained by chiral
 symmetry. At low momentum they are described by an effective Lagrangian
that transforms correctly with respect to chiral $SU(3)_L\times SU(3)_R$
symmetry and has the least number of derivatives or insertions of the quark
mass matrix
$$m_q=\pmatrix{m_u&0&0\cr
               0&m_d&0\cr
               0&0&m_s\cr}\,\, . \eqno (9)$$
The pseudo-Goldstone bosons $\pi$, $K$ and $\eta$ are incorporated in a
$3\times 3$ special unitary matrix
$$\Sigma ={\rm exp}(2iM/f)\,\, , \eqno (10)$$
where
$$M=\pmatrix{\pi ^0/\sqrt {2}+\eta /\sqrt {6}&\pi ^+&K^+\cr
              \pi ^-&-\pi ^0 /\sqrt {2}+\eta /\sqrt {6}&K^0\cr
              K^-&\bar K^0&-2\eta /\sqrt {6}\cr} \,\, . \eqno (11)$$
The matrix $\Sigma $ transforms under chiral $SU(3)_L\times SU(3)_R$ as
$$\Sigma \rightarrow L\Sigma R^\dagger \,\, , \eqno (12)$$
where $L$ is a member of $SU(3)_L$ and $R$ is a member of $SU(3)_R$.
The strong interactions of the $\pi$, $K$ and $\eta $ are described by the
effective Lagrangian density
$${\cal L}= {f^2\over 8}{\rm Tr}\partial_{\mu}\Sigma\partial^{\mu}\Sigma
      ^{\dagger}+v{\rm Tr}(m_q\Sigma + \Sigma^{\dagger}m_q) \,\, ,\eqno (13)$$
where $f\simeq 132 ~{\rm MeV}$ is the pion decay constant. The first term
in eq. (13) is invariant under chiral $SU(3)_L \times SU(3)_R$ and the
second term transforms as $(3_L,\bar 3_R)+(\bar 3_L,3_R)$. The enhanced part
of the effective Hamiltonian for weak $\Delta S=1$ nonleptonic kaon decays
transforms as $(8_L,1_R)$ under chiral symmetry and is given by
$${\cal H}^{|\Delta S=1|}={g_8G_Ff^4\over 4\sqrt{2}}V_{us}^*V_
{ud}{\rm Tr}O_W\partial _{\mu}\Sigma \partial ^{\mu}\Sigma ^{\dagger} + h.c.
\,\, , \eqno (14)$$
where
$$O_W=\pmatrix{0&0&0\cr
               0&0&0\cr
               0&1&0\cr}\,\, , \eqno (15)$$
projects out the correct part of the octet and the measured $K_S \rightarrow
 \pi ^+\pi ^-$ decay amplitude determines that $|g_8|\simeq 5.1$.

$J_{8\mu}^{(L)}$ and $J_{\mu}^{(e.m.)}$ are currents associated with generators
 of chiral symmetry. At leading order in chiral perturbation theory the
$K^+\rightarrow \pi ^+Z^0$ vertex that arises from the $Z^0$ coupling to these
currents is obtained by gauging the  chiral Lagrangian density (13) and the
Hamiltonian density (14) via the replacement
$$\partial _{\mu}\Sigma \rightarrow D_{\mu}\Sigma = \partial _{\mu}\Sigma
+i\sqrt{g_1^2+g_2^2}Z_{\mu}^0~(Q\Sigma -{\rm sin^2\theta _W}[Q,\Sigma ])
\,\, , \eqno (16)$$
where
$$Q=\pmatrix{{2\over 3}&0&0\cr
             0&-{1\over 3}&0\cr
              0&0&-{1 \over 3}\cr}\,\,  \eqno (17)$$
is the electromagnetic charge matrix,
and computing the tree level Feynman diagrams in Fig. (1) (using the
interactions that follow from eqs. (13)-(17)). In Fig. (1) the incoming
dashed line denotes the $K^+$, the outgoing dashed line denotes the
$\pi^+$ and the wiggly line denotes the $Z^0$.  (Only some of the diagrams in
Fig. 1 were considered by Ref. [5].)  We are interested
in the part of the $K^+\rightarrow \pi^+ Z^0$ vertex proportional
to $(p_K+p_{\pi})^{\mu}$. The part proportional to $(p_K-p_{\pi})^{\mu}$
doesn't contribute to $K^+\rightarrow \pi ^+ \nu\bar \nu$ since the neutrinos
are massless. The $Z^0$ coupling to $J_{\mu}^{(e.m.)}$ doesn't give rise
to a $K^+\rightarrow \pi ^+ Z^0$ vertex at the leading order of chiral
perturbation theory (the three diagrams in Fig. (1)
cancel [6]).  The $Z^0$  coupling to $J_{8\mu}^{(L)}$ gives
$$ {G_F\over 2\sqrt 2}\sqrt {g_1^2+g_2^2}~g_8V_{us}^*V_{ud}f^2
\left [0+1-{2\over 3}\right](p_K+p_{\pi})^{\mu} \,\, , \eqno (18)$$
for the $K^+\rightarrow \pi ^+ Z^0$ vertex ($m_u$  and $m_d$ are neglected
here). The three terms $0, 1$ and $- 2/3$ in the
square brackets of eq. (18) arise from Figs. (1a), (1b) and (1c)
respectively. The long distance contribution to $K^+\rightarrow \pi ^+ \nu\bar
 \nu$ that results from the vertex in eq. (18) is
$$ \xi_{LD}^{(8)}={g_8\pi ^2\over 3}(f/M_W)^2 \,\, . \eqno (19)$$
Numerically eq. (19) is about $5 \times 10^{-5}$ which is only
 $5\% $ of the charm quark short distance contribution, $\xi_c$.

The $K^+\rightarrow \pi ^+ Z^0$ vertex arising form the $Z^0$ coupling
to $J_{1\mu}^{(L)}$ cannot be calculated using chiral perturbation theory
 alone because through the anomaly
instanton field configurations in the QCD path integral
 spoil the axial $U(1)$ symmetry. (It is expected to have the same order of
magnitude as eq. (18).) However, in the large $N_c$ limit [7]
effects of the anomaly
 are suppressed and the axial $U(1)$ is a good symmetry [8].
Then the $Z^0$ coupling to $J_{1\mu}^{(L)}$ is taken into account by
adding to the covariant derivative in eq. (16) the term
$$i\sqrt {g_1^2+g_2^2}Z_{\mu}^0~(-{1\over 6}\Sigma)\,\, . \eqno (20)$$
In the large $N_c$ limit the coupling of the $Z^0$ to $J_{1\mu}^{(L)}$
gives rise to the $K^+\rightarrow \pi ^+ Z^0$ vertex
$${G_F\over 2\sqrt{2}}\sqrt {g_1^2+g_2^2}~g_8V_{us}^*V_{ud}f^2
\left[0+0-{1\over 3}\right](p_K+p_{\pi})^{\mu}\,\, .\eqno (21)$$
The three terms $0,0$ and $-1/3$ in the square brackets of eq. (21) come
respectively from
Figs. (1a), (1b) and (1c). This implies that the $Z^0$ coupling to
$J_{1\mu}^{(L)}$ gives a contribution to $\xi_{LD}$ that satisfies
$$\lim_{N_c\to\infty}\xi_{LD}^{(1)}=-{g_8\pi ^2\over 3}(f/M_W)^2\,\, . \eqno
 (22)$$
This cancels the contribution in eq. (19).

In the large $N_c$ limit the leading (in chiral perturbation theory)
long distance contribution to $K^+\rightarrow \pi ^+\nu \bar \nu$ from
Feynman diagrams with a $W$ boson and a $Z^0$ boson vanishes. However, in the
large
 $N_c$ limit the $\eta'$ is a pseudo-Goldstone boson ($m_{\eta^\prime}^2$
is of order $1/N_c$). It's large mass,
$958$  MeV, is an indication that this limit is not trustworthy [9].
Nonetheless, it seems likely that some remnant of the cancellation that occurs
as $N_c \rightarrow \infty$ survives in the physical case of three
colors. We do not expect the long distance contribution to
$K^+\rightarrow \pi ^+\nu \bar \nu$ from Feynman diagrams with a
$W$ and $Z^0$ to exceed eq. (19). Contributions from Feynman diagrams
with two $W$ bosons are expected to be even smaller. They are not
enhanced by the factor $g_8$ which reflects the $\Delta I=1/2$
rule.
\bigskip

\centerline {\bf References}

\item{1.} T. Inami and C.S. Lim, Prog. Theor. Phys. {\bf 65} (1981)
297;1772.

\item{2.} G. Buchalla and A.J. Buras, MPI-Ph/93-44 (unpublished).

\item{3.} J. Ellis and J.S. Hagelin, Nucl. Phys. {\bf B217} (1983) 189;
V.A. Novikov, M. A. Shifman, A.I. Vainshtein and V.I. Zakharov,
Phys. Rev. {\bf D16} (1977) 223; C.O. Dib, I.Dunietz and F.J. Gilman,
Mod. Phys.  Lett. {\bf A6} (1991) 3573; C. Buchalla, A.J. Buras and M.K.
Harlander Nucl. Phys. {\bf B349} (1991) 1.

\item{4.} D. Rein and L.M. Sehgal, Phys. Rev. {\bf D39} (1989) 3325.

\item{5.}  J.S. Hagelin and L.S. Littenberg, Prog. Part. Nucl. Phys. {\bf 23}
(1989) 1.

\item{6.} G. Ecker, A. Pich, E. De Rafael, Nucl Phys. {\bf B303}
 (1988) 665.

\item{7.} G. 't Hooft, Nucl. Phys. {\bf B72} (1974) 461.

\item{8.} E. Witten, Nucl. Phys. {\bf B156} (1979) 269; P. Di Vecchia
and G. Veneziano, Nucl. Phys. {\bf B171} (1980) 253.

\item{9.} H. Georgi, HUTP-93-A029 (1993) (unpublished).
\bigskip

\centerline {\bf Figure Caption}

Feynman diagrams that dominate the $K^+ \rightarrow \pi^+ Z^0$ vertex in chiral
perturbation theory.  The shaded circle denotes an insertion of the weak
$\Delta S = 1$ nonleptonic Hamiltonian in eq. (14).
\bye